\documentclass[aps,english,preprint,nofootinbib,preprintnumbers,superscriptaddress]{revtex4}
\usepackage{mathrsfs}
\usepackage{hyperref}
\hypersetup{
    colorlinks=true,
    linkcolor=blue,
    filecolor=magenta,      
    urlcolor=cyan,
    pdftitle={Overleaf Example},
    pdfpagemode=FullScreen,
    }
\usepackage{graphicx}
\usepackage{amsmath, amssymb}
\usepackage{babel}
\usepackage{color}
\usepackage{slashed}
\usepackage[T1]{fontenc}
\usepackage{titlesec}
\def\beqn{\begin{eqnarray}}
\def\eeqn{\end{eqnarray}}
\def\barr{\begin{array}}
\def\earr{\end{array}}
\def\btab{\begin{tabular}}
\def\etab{\end{tabular}}
\def\bite{\begin{itemize}}
\def\eite{\end{itemize}}
\def\bcen{\begin{center}}
\def\ecen{\end{center}}

\def\ba#1\ea{\begin{align}#1\end{align}}
\newcommand{\nn}{\nonumber}
\newcommand{\ra}{\rangle}
\newcommand{\la}{\langle}

\begin{document}
%\preprint{MITP/20-019}

\title{Quark mass difference effects in hadronic Fermi matrix elements from first principles}

\newcommand\uconn{Physics Department, University of Connecticut, Storrs, CT 06269-3046, USA}
\newcommand\uw{Department of Physics, University of Washington, Seattle, WA 98195-1560, USA}

\author{Chien-Yeah Seng}

\affiliation{Facility for Rare Isotope Beams, Michigan State University, East Lansing, MI 48824, USA}
\affiliation{\uw}

\author{Vincenzo Cirigliano}
\affiliation{Institute for Nuclear Theory, University of Washington, Seattle, WA  98195-1550}

\author{Xu Feng}
\affiliation{School of Physics, Peking University, Beijing 100871, China}
\affiliation{Collaborative Innovation Center of Quantum Matter, Beijing 100871, China}
\affiliation{Center for High Energy Physics, Peking University, Beijing 100871, China}

\author{Mikhail Gorchtein}
\affiliation{
Institut f\"ur Kernphysik, Johannes Gutenberg-Universit\"{a}t,
	J.J. Becher-Weg 45, 55128 Mainz, Germany}
 \affiliation{
PRISMA+ Cluster of Excellence, Johannes Gutenberg-Universit\"{a}t, Mainz, Germany 
}

\author{Luchang~Jin}
\affiliation{\uconn}

\author{Gerald A. Miller}
\affiliation{\uw}

\date{\today}

\begin{abstract}

It was recently estimated that the strong isospin-symmetry breaking (ISB) corrections to the Fermi matrix element in free neutron decay could be of the order $10^{-4}$, one order of magnitude larger than the na\"{\i}ve estimate based on the Behrends-Sirlin-Ademollo-Gatto theorem. To investigate this claim, we derive a general expression of the leading ISB correction to hadronic Fermi matrix elements, which takes the form of a four-point correlation function in lattice gauge theory and is  straightforward to compute from first principles. Our formalism paves the way for the first  determination of such correction in the neutron sector with fully-controlled theory uncertainties.  

\end{abstract}

\maketitle

%  \tableofcontents

\section{Introduction}

The strength of the Fermi transition in beta decays of hadrons and nuclei at tree level is determined by the Fermi matrix element $M_F$, namely the matrix element of the isospin-raising (or lowering) operator between the initial and final states. In the isospin-symmetric limit, $M_F$ is fully determined by group theory, but the presence of isospin-symmetry-breaking (ISB) effect induces a small correction that needs to be taken into account for precision tests of the Standard Model (SM). In particular, the Cabibbo-Kobayashi-Maskawa (CKM) matrix element $V_{ud}$ extracted from neutron and nuclear beta decays is approaching a $10^{-4}$ precision~\cite{Zyla:2020zbs}, which calls for the same precision level for all SM theory inputs to the decay processes.   

According to the well-known theorem by Behrends \& Sirlin  and  Ademollo \& Gatto (BSAG)~\cite{Behrends:1960nf,Ademollo:1964sr} (see also Ref.\cite{terent1963conservation}), the deviation of $M_F$ from its isospin-symmetric limit starts at second order in the ISB interactions. In nuclear beta decays where the primary source of ISB is the Coulomb repulsion between protons that scales as $Z\alpha$, such correction is at (0.1-1)\% level which is substantial. On the other hand, there is no large $Z$-enhancement in the hadron sector (e.g. pions and neutron), so one expects both strong and electromagnetic ISB corrections to take the natural size. For instance, an explicit calculation using Chiral Perturbation Theory (ChPT) indicates that the ISB effect in pion beta decay is below $10^{-5}$ which can be very safely neglected~\cite{Cirigliano:2002ng}.

The situation is more complicated for neutron decay. While it is widely accepted that the electromagnetically induced ISB effect is small, some recent analyses based on quark models suggested that the strong ISB effect could be on the order of $10^{-4}$~\cite{Guichon:2011gc,Crawford:2022yhi}. This exceeds earlier estimates~\cite{Donoghue:1990ti} based on the scaling  $(m_u-m_d)^2/\Lambda_{QCD}^2\sim 10^{-5}$, due to the coherent enhancement from the various excited state contributions which avoids the cancellation between electromagnetic and quark mass difference effects present in computing the mass difference. Given that a $10^{-4}$ correction is relevant to precision experiments,
it is highly desirable to check this assertion using first-principles calculations with lattice Quantum Chromodynamics (QCD). A conventional way to study ISB corrections is to set up the calculation in the isospin-symmetric limit and treat the quark mass splitting as a perturbation. Observables such as the hadron mass splitting and the Dashen's theorem breaking
parameter that scale linearly to the ISB parameters have been studied with this method~\cite{deDivitiis:2013xla}, but the correction to $M_F$, which is a quadratic effect, remains unexplored. 

In this short paper we outline a procedure that allows for straightforward lattice calculations of strong ISB corrections to hadronic Fermi matrix elements to satisfactory precision. It is based on a method similar to the one   developed earlier by some of us for nuclear beta decays~\cite{Seng:2023cvt}, but without any of the model assumptions present in the latter. In short, one computes the leading strong ISB correction derived from perturbation theory instead of the full Fermi matrix element; the former takes the form of a hadronic matrix element of two isotriplet scalar operators sandwiching a Green's function $1/(H_0-\zeta)$ in a fully isospin-symmetric system. One then re-expresses the matrix element in terms of a four-point correlation function which is readily calculable on lattice.  Given the smallness of the  ISB pre-factor, a $\sim$20\% accuracy for such a matrix element is sufficient. We focus in this paper mainly on the theory formalism, and defer the actual lattice calculation to a future work. 

\section{Fermi matrix element}

Let us consider a generic allowed $\beta^-$-decay $\phi_i\rightarrow \phi_f e\bar{\nu}_e$ which is triggered by the following charged weak current:
\begin{equation}
J_W^\mu(x)=\bar{u}(x)\gamma^\mu(1-\gamma_5)d(x)~.
\end{equation}
The spatial integral of the zeroth component of the vector piece defines the isospin-raising operator:
\begin{equation}
\hat{\tau}_+=\int d^3xu^\dagger(\vec{x})d(\vec{x})=\int d^3x (J_W^0(\vec{x}))_V~.
\end{equation} 
Here, we adopt the particle physics' convention of isospin, i.e. $T_z(u)=+1/2$. 

In quantum field theory, plane-wave states are normalized to a delta function:
\begin{equation}
\langle \phi(\vec{p}_f,s_f)|\phi(\vec{p}_i,s_i)\rangle=(2\pi)^3 2E(\vec{p}_i)\delta^{(3)}(\vec{p}_i-\vec{p}_f)\delta_{s_i,s_f}~,
\end{equation}
which means zero-momentum states are normalized as:
\begin{equation}
\langle \phi(\vec{0},s_f)|\phi(\vec{0},s_i)\rangle\rightarrow(2\pi)^3 2m\delta^{(3)}(\vec{0})\delta_{s_i,s_f}=2mL^3\delta_{s_i,s_f}~,
\end{equation}  
if we restrict ourselves to a finite box of the size $L$. We may define a quantum-mechanical state $|\phi(s)\rangle$ from a zero-momentum plane-wave state
\begin{equation}
|\phi(s)\rangle\equiv \frac{1}{\sqrt{2mL^3}}|\phi(\vec{0},s)\rangle~,
\end{equation}
normalized as 
%$\langle \phi_{s'}|\phi_{s}\rangle=\delta_{s,s'}$. 
$\langle \phi (s')|\phi (s)\rangle=\delta_{s,s'}$. 
With this, the Fermi matrix element in a $\beta^-$-decay may be  defined as:
\begin{equation}
M_F\equiv \langle \phi_f(s)|\hat{\tau}_+|\phi_i(s)\rangle~.\label{eq:MF}
\end{equation} 
Notice that, according to this definition the Fermi matrix element involves states with vanishing three-momentum and  is related to hadronic form factors with four-momentum transfer  $q^2=(m_f-m_i)^2$, rather than  $q^2=0$. 

\section{Perturbation theory and ISB corrections}

To study ISB effects, we split the full Hamiltonian into:
\begin{equation}
H=H_0+V~,
\end{equation}
where $H_0$ is the isospin-symmetric part and $V$ is the ISB perturbation term. The states $|\phi_{i,f}(s)\rangle$ in Eq.\eqref{eq:MF} are eigenstates of the full Hamiltonian $H$ while the corresponding eigenstates of $H_0$ are denoted as $|\phi_{i,f}(s)\rangle_0$ with degenerate mass eigenvalue $m_\phi^0$. They are also exact isospin eigenstates within the same isomultiplet, connected through the isospin-raising operator as:
\begin{equation}
M_F^0|\phi_f(s)\rangle_0=\hat{\tau}_+|\phi_i(s)\rangle_0~,
\end{equation}
where $M_F^0$ is the bare Fermi matrix element.
 
Following the notation in nuclear beta decay, we define the ISB correction $\delta$ to the Fermi matrix element as:
\begin{equation}
M_F^2=(M_F^0)^2(1-\delta)~.\label{eq:MFdev}
\end{equation}
An exact expression of $\delta$ can be derived using the Brillouin-Wigner perturbation theory in Quantum Mechanics, which was first adopted in Refs.\cite{Miller:2008my,Miller:2009cg} for nuclear beta decays. 
Consistently with the BSAG theorem~\cite{Behrends:1960nf,Ademollo:1964sr}, one finds that the leading order correction scales as $\mathcal{O}(V^2)$~\cite{Seng:2023cvt}
\begin{eqnarray}
\delta &\approx &{}_0\langle\phi_i(s)|V\Lambda_i\left(\frac{1}{m_\phi^0-\Lambda_i H_0\Lambda_i}\right)^2\Lambda_i V|\phi_i(s)\rangle_0\nonumber\\
&&+\:{}_0\langle\phi_f(s)|V\Lambda_f\left(\frac{1}{m_\phi^0-\Lambda_f H_0\Lambda_f}\right)^2\Lambda_f V|\phi_f(s)\rangle_0\nonumber\\
&&-\frac{2}{M_F^0}\:{}_0\langle \phi_f(s)|V\Lambda_f\frac{1}{m_\phi^0-\Lambda_f H_0\Lambda_f}\hat{\tau}_+\frac{1}{m_\phi^0-\Lambda_i H_0\Lambda_i}\Lambda_iV|\phi_i(s)\rangle_0~,\label{eq:delta}
\end{eqnarray}
where $\Lambda_{i,f}\equiv 1-|\phi_{i,f}(s)\rangle_0{}_0\langle \phi_{i,f}(s)|$ is an operator that projects away the unperturbed state $|\phi_{i,f}(s)\rangle_0$. Eq.\eqref{eq:delta} serves as the foundation of our further analysis.

In this work we concentrate on the strong ISB effects, whose only source is the $u-d$ quark mass difference term in the QCD Lagrangian which is purely isovector:
\begin{equation}
V=-\int d^3x\mathcal{L}_{u-d}\equiv \frac{\Delta m_q}{2}\hat{O}^1_0~,
\end{equation}
where $\Delta m_q=m_u-m_d$ is the quark mass splitting. We introduce the following rank-one tensor operators in  isospin space for future convenience:
\begin{equation}
\hat{O}_0^1=\int d^3x(\bar{u}(\vec{x})u(\vec{x})-\bar{d}(\vec{x})d(\vec{x}))~,~
\hat{O}_{-1}^1=\sqrt{2}\int d^3x\bar{d}(\vec{x})u(\vec{x})~,~\hat{O}_{+1}^1=-(\hat{O}_{-1}^1)^\dagger~.
\label{eq:Oi}
\end{equation}
With the aid of these operators, we can simplify the right hand side of Eq.\eqref{eq:delta} into more elegant expressions. We will do this for pion and neutron decays.

\subsection{$\pi^-\rightarrow \pi^0 e\bar{\nu}_e$}

As a first example we consider  pion beta decay. As isospin eigenstates, we can label $|\pi_0\rangle_0=|\pi;1,0\rangle$, $|\pi_+\rangle_0=-|\pi;1,+1\rangle$ and $|\pi_-\rangle_0=|\pi;1,-1\rangle$, with the bare Fermi matrix element $M_F^0={}_0\langle \pi_0|\hat{\tau}_+|\pi_-\rangle_0=\sqrt{2}$. We start from Eq.\eqref{eq:delta} and insert a complete set of $H_0$ eigenstates $\{|a;T,T_z\rangle\}$ with energy $E_{a,T}$, where $a$ denotes all quantum numbers unrelated to isospin (the single-pion states are automatically excluded by the projection operators). Due to the isovector nature of $V$,  only $T=0,1,2$ intermediate states contribute. Applying the Wigner-Eckart theorem
\begin{equation}
\langle a;T,T_z|\hat{O}^1_{T_z''}|\pi;1,T_z'\rangle=C_{1,T_z';1,T_z''}^{T,T_z}\langle a;T||\hat{O}^1||\pi\rangle
\end{equation}
where $C_{1,T_z';1,T_z''}^{T,T_z}$ are Clebsch-Gordan coefficients and $\langle a;T||\hat{O}^1||\pi\rangle$ is a reduced matrix element, we  obtain:
\begin{equation}
\delta_\pi=\frac{(\Delta m_q)^2}{4}\left[\frac{1}{3}\sum_a\frac{|\langle a;0||\hat{O}^1||\pi\rangle|^2}{(m_\pi^0-E_{a,0})^2}+\frac{1}{2}\sum_a\frac{|\langle a;1||\hat{O}^1||\pi\rangle|^2}{(m_\pi^0-E_{a,1})^2}-\frac{5}{6}\sum_a\frac{|\langle a;2||\hat{O}^1||\pi\rangle|^2}{(m_\pi^0-E_{a,2})^2}\right]~,
\label{eq:deltapion}
\end{equation}
analogous to the result in Ref.\cite{Seng:2023cvt}. 

To compute $\delta_\pi$, we define a ``generating function'' $\mathcal{F}_\pi(\zeta)$ which involves diagonal matrix elements of two $\hat{O}_i^1$ operators and a Green's function of the isospin-symmetric system:
\begin{equation}
\mathcal{F}_\pi(\zeta)\equiv \langle \pi;1,T_z|(\hat{O}_{-1}^1)^\dagger \frac{1-\hat{P}_\pi}{H_0-\zeta}\hat{O}_{-1}^1|\pi;1,T_z\rangle~,
\end{equation}
where $\zeta$ is a free energy parameter, and 
\ba
\hat P_\pi
=
\sum_{T_z}\int \frac{d^3 p}{(2\pi)^32E_{\pi}(\vec{p})} | \pi (\vec p);1,T_z \ra  \la \pi (\vec p);1,T_z | \label{eq:pionproject}
\ea
is a projection operator of the single-particle pion isotriplet states (the subscript $\pi$ in $\mathcal{F}_\pi$, which can take $\pi^{\pm}$ or $\pi^0$, affects only the external state but not the projection operator), so $1-\hat{P}_\pi$ projects away such states. One may check that the combination
\begin{equation}
\mathcal{F}_{\pi^+}(\zeta)-\mathcal{F}_{\pi^-}(\zeta)=-\frac{1}{3}\sum_a\frac{|\langle a;0||\hat{O}^1||\pi\rangle|^2}{\zeta-E_{a,0}}-\frac{1}{2}\sum_a\frac{|\langle a;1||\hat{O}^1||\pi\rangle|^2}{\zeta-E_{a,1}}+\frac{5}{6}\sum_a\frac{|\langle a;2||\hat{O}^1||\pi\rangle|^2}{\zeta-E_{a,2}}
\end{equation}
involves the same relative coefficients between reduced matrix elements of different $T$
as in the correction $\delta$, see Eq.~(\ref{eq:deltapion}). Therefore $\delta$ is simply obtained by taking the derivative at $\zeta=m_\pi^0$: 
\begin{equation}
\delta_\pi=\frac{(\Delta m_q)^2}{4}(\mathcal{F}'_{\pi^+}(m_\pi^0)-\mathcal{F}'_{\pi^-}(m_\pi^0))~.
\end{equation}

\subsection{$n\rightarrow p e\bar{\nu}_e$}

We work out the same derivation for neutron decay. In the isospin limit we label the nucleon doublet as $|p(s)\rangle_0=|N(s);1/2,+1/2\rangle$ and $|n(s)\rangle_0=|N(s);1/2,-1/2\rangle$, with the bare Fermi matrix element $M_F^0={}_0\langle p(s)|\hat{\tau}_+|n(s)\rangle_0=1$. A complete set of $H_0$ eigenstates $\{|a;T,T_z\rangle\}$ excluding the ground-state nucleon doublet is introduced, and only $T=1/2,3/2$ states are relevant. Using the Wigner-Eckart theorem we obtain:
\begin{equation}
\delta_N=\frac{(\Delta m_q)^2}{3}\left[\sum_a\frac{|\langle a;1/2||\hat{O}^1||N(s)\rangle|^2}{(m_N^0-E_{a,1/2})^2}-\sum_a\frac{|\langle a;3/2||\hat{O}^1||N(s)\rangle|^2}{(m_N^0-E_{a,3/2})^2}\right]~.
\end{equation}
Since the expression above involves a sum over all contributing excited states $a$, a coherent addition could lead to an unexpected enhancement over the na\"{\i}ve power counting result. This was the main speculation in Refs.\cite{Guichon:2011gc,Crawford:2022yhi}, which can now be tested directly on the lattice.
The generating function is defined similarly as: 
\begin{equation}
\mathcal{F}_N(\zeta)\equiv \langle N(s)|(\hat{O}_{-1}^1)^\dagger \frac{1-\hat{P}_N}{H_0-\zeta}\hat{O}_{-1}^1|N(s)\rangle~,
\end{equation}
and one may verify that
\begin{equation}
\mathcal{F}_{p}(\zeta)-\mathcal{F}_{n}(\zeta)=-\frac{2}{3}\sum_a\frac{|\langle a;1/2||\hat{O}^1||N(s)\rangle|^2}{\zeta-E_{a,1/2}}+\frac{2}{3}\sum_a\frac{|\langle a;3/2||\hat{O}^1||N(s)\rangle|^2}{\zeta-E_{a,3/2}}~,
\end{equation}
which implies
\begin{equation}
\delta_N=\frac{(\Delta m_q)^2}{2}(\mathcal{F}'_{p}(m_N^0)-\mathcal{F}'_{n}(m_N^0))~.
\end{equation}

An obvious advantage of this formalism is that one focuses directly on the small quantity $\delta_N$ instead of the full $M_F$, which significantly reduces the level of theoretical precision needed. Suppose we consider  the estimate of Ref.\cite{Crawford:2022yhi} where $\delta_N\sim 4\times 10^{-4}$ for neutron decay (after averaging over three models), then a $\sim 20$\% precision level for the non-perturbative calculation of $\mathcal{F}_N'(m_N^0)$ would be sufficient to keep the theory uncertainty of $M_F^2$ below $10^{-4}$. This is in clear contrast to the direct calculation of $M_F$ which would require a $10^{-4}$ precision. 

\section{Lattice implementation}

\begin{figure}
	\centering
	\includegraphics[width=0.7\columnwidth]{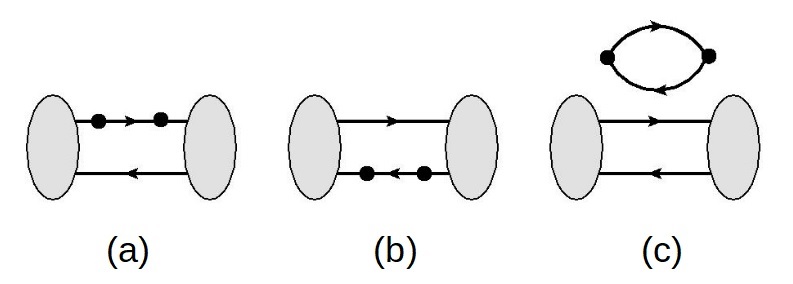}
	\caption{Quark contraction diagrams involved in $C_\pi(t)$: Grey blobs denote external pion states and the black dots denote insertions of the operators $\hat{O}_{1,2}$, where 
 $\hat{O}_1=(\hat{O}_{-1}^1)^\dagger$, $\hat{O}_2=\hat{O}_{-1}^1$, with $\hat{O}_{-1}^1$ defined in Eq.~(\ref{eq:Oi}).}
	\label{fig:contraction}
\end{figure}

Here we briefly discuss how the aforementioned generating function and its derivative are implemented on the lattice, taking the pion as an example. We define the following four-point correlation function ($t \geq 0$) in the isospin-symmetric limit, which can be directly calculated with lattice QCD:
\ba
\label{eq:pion-projection-in-corr}
C_\pi(t)
&=
\la \pi | \hat{O}_1(t) \hat{O}_2(t=0) | \pi \ra
-\la \pi | \hat{O}_1(t) \hat P_\pi \hat{O}_2(t=0) | \pi \ra
\\
&=
\la \pi | \hat{O}_1(t) (1 - \hat P_\pi) \hat{O}_2(t=0) | \pi \ra
\nn\\
&=
\la \pi | \hat{O}_1(t=0) e^{-(H_0 - m_\pi^0 )t} (1 - \hat P_\pi) \hat{O}_2(t=0) | \pi \ra
\ea
where $\hat{O}_1=(\hat{O}_{-1}^1)^\dagger$, $\hat{O}_2=\hat{O}_{-1}^1$, and the time dependence of the operator comes from the standard Euclidean space-time Heisenberg picture:
\ba
\hat{O}(t) = e^{H_0t} \hat{O}(t=0) e^{-H_0t}~.
\ea
The correlation function involves quark contraction diagrams as depicted in Fig.\ref{fig:contraction}.
The matrix elements in Eq.~\eqref{eq:pion-projection-in-corr} that involve the zero-momentum pion projection operator $\hat P_\pi$ (see Eq.\eqref{eq:pionproject}) can be calculated as the product of two simpler matrix elements which require a separate calculation of the lattice three-point functions with pion initial/final states and a single operator $\hat O_{1/2}$.

Both the generating function $\mathcal{F}_\pi(\zeta)$ and its derivative at $\zeta=m_\pi^0$ can now be obtained as:
\ba
\mathcal{F}_\pi(m_\pi^0)=\int_0^\infty  C_\pi(t)\, dt
=
\la \pi | \hat{O}_1(t=0) \frac{1 - \hat P_\pi}{H_0 - m_\pi^0} \hat{O}_2(t=0) | \pi \ra
\ea
and
\ba
\mathcal{F}_\pi'(m_\pi^0)=\int_0^\infty t C_\pi(t)\, dt
=
\la \pi | \hat{O}_1(t=0) \frac{1 - \hat P_\pi}{(H_0 - m_\pi^0)^2}  \hat{O}_2(t=0) | \pi \ra
\ea
respectively. The same also works for the nucleon. 

One could argue that a 20\% precision on a four-point correlation function is not necessarily easier than a $10^{-4}$ precision on the direct calculation of $M_F$; in the latter, one may compute the derivative in the ISB parameter numerically and exploit the correlation between the calculation performed with different values of the ISB, which has been proven successful in the past studies of meson mass splittings. However, the fact that the ISB correction to $M_F$ is a quadratic effect (in contrast to the linear correction to masses) may complicate the procedure. Furthermore, within our theoretical framework, one may also compute the four-point function implicitly with numerical derivatives of two-point meson correlation functions with respect to the quark masses. This method is also commonly used in many lattice calculations with comparable precision as calculating the four-point correlation function directly~\cite{Feng:2021zek,Frezzotti:2022dwn,Blum:2007cy,BMW:2014pzb}.

\section{Insights from ChPT}

It is useful to gauge first-principles studies of $\delta$ with known results in the pion sector. We start by connecting our definition of $M_F$ to the relativistic charged weak form factors:
\begin{equation}
\langle \pi_0(p_f)|J_W^\mu(0)|\pi_-(p_i)\rangle=f_+(q^2)(p_i+p_f)^\mu+f_-(q^2)(p_i-p_f)^\mu~,
\end{equation} 
where
\begin{equation}
f_+(0)\rightarrow M_F^0=\sqrt{2}~,~f_-(q^2)\rightarrow 0~
\end{equation}
in the isospin limit.
Following common practice, we scale out the constant factor $f_+(0)$:
\begin{equation}
\bar{f}_{\pm}(q^2)=\frac{1}{f_+(0)}f_{\pm}(q^2)~.
\end{equation}
Also, for the form factor $\bar{f}_+(q^2)$ we need its leading $q^2$-dependence, which defines a mean square radius $\langle r_W^2\rangle$:
\begin{equation}
\bar{f}_+(q^2)=1+\frac{q^2}{6}\langle r_W^2\rangle +\mathcal{O}(q^4)~.
\end{equation}
Then, by choosing $\vec{p}_i=\vec{p}_f=\vec{0}$ as we advocated above, the full Fermi matrix element can be expressed in terms of the form factors as:
\begin{equation}
M_F=f_+(0)\left[\frac{m_{\pi_-}+m_{\pi_0}}{2\sqrt{m_{\pi_-}m_{\pi_0}}}\bar{f}_+\left((m_{\pi_-}-m_{\pi_0})^2\right)+\frac{m_{\pi_-}-m_{\pi_0}}{2\sqrt{m_{\pi_-}m_{\pi_0}}}\bar{f}_-\left((m_{\pi_-}-m_{\pi_0})^2\right)\right]~.\label{eq:MFpion}
\end{equation}
For future convenience, we also define the average pion mass and the pion mass splitting as:
\begin{equation}
\bar{m}_\pi\equiv \frac{m_{\pi_-}+m_{\pi_0}}{2}~,~\Delta m_\pi\equiv m_{\pi_-}-m_{\pi_0}~.
\end{equation}

By parameterizing the deviation of $f_+^2(0)$ from its isospin limit as:
\begin{equation}
f_+^2(0)\equiv(M_F^0)^2(1-\delta_{f_+})
\end{equation}
where $\delta_{f_+}=\mathcal{O}(V^2)$, we can expand Eq.\eqref{eq:MFpion} to the leading order in strong ISB parameters, which provides a parameterization of $\delta_\pi$:
\begin{equation}
\delta_\pi\approx \delta_{f_+}-\frac{(\Delta m_\pi)^2}{4\bar{m}_\pi^2}-\frac{(\Delta m_\pi)^2}{3}\langle r_W^2\rangle-\frac{\Delta m_\pi}{\bar{m}_\pi}\bar{f}_-(0)~.
\label{da}\end{equation}
The four terms at the right hand side represent (1) The deviation of $f_+(0)$ from isospin limit, (2) The correction to the external state normalization, (3) The finite-$q^2$ correction, and (4) The effect of the subdominant form factor $f_-$, respectively.

Predictions of  the four terms are available
using the three-flavor ChPT~\cite{Gasser:1984gg}. To leading chiral order, the pseudoscalar meson octet masses are given in terms of quark masses by 
\begin{eqnarray}
m_{\pi_+}^2&=&2B_0\hat{m}\nonumber\\
m_{\pi_0}^2&=&2B_0\hat{m}-\frac{B_0\hat{m}^2\varepsilon^2}{m_s-\hat{m}}\nonumber\\
m_{K_+}^2&=&B_0(m_s+\hat{m}(1-\varepsilon))\nonumber\\
m_{K_0}^2&=&B_0(m_s+\hat{m}(1+\varepsilon))\nonumber\\
m_{\eta_0}^2&=&\frac{2}{3}B_0(2m_s+\hat{m})+\frac{B_0\hat{m}^2\varepsilon^2}{m_s-\hat{m}}~,
\end{eqnarray}
where $\hat{m}=(m_u+m_d)/2$, $\varepsilon=(m_d-m_u)/(m_u+m_d)$, and the constant $B_0$ characterizes the strength of the chiral condensate. The $\pi_+-\pi_0$ mass splitting induced by the $\pi_3-\eta_8$ mixing is quadratic to $m_u-m_d$. This means  the last three terms at the right hand side of Eq.\eqref{da} scale as $\mathcal{O}((m_u-m_d)^3)$ or higher, and therefore we drop them. This leaves us with $\delta_{f_+}$, which we take from \cite{Cirigliano:2002ng}:
\begin{equation}
\delta_{f_+}\approx -4H_{\pi_+\pi_0}(0)-2H_{K_+K_0}(0)~,~H_{PQ}(0)\approx-\frac{1}{192\pi^2}\frac{(m_P^2-m_Q^2)^2}{F_0^2(m_P^2+m_Q^2)}~,
\end{equation}
where $F_0$ is the pion decay constant in the chiral limit. The only contribution quadratic to $m_u-m_d$ comes from the kaon loop function $H_{K_+ K_0}$, which gives:
\begin{equation}
\delta_\pi\approx\delta_{f_+}\approx\frac{B_0(m_u-m_d)^2}{192\pi^2F_0^2(m_s+\hat{m})}~.
\end{equation}
We take the QCD parameters from the most recent FLAG review in the $\overline{\text{MS}}$ scheme, at the renormalization scale $\mu=2$~GeV with $N_f=2+1$~\cite{FlavourLatticeAveragingGroupFLAG:2021npn}: $F_0=80.3(6.0)$~MeV~\cite{MILC:2010hzw}, $\Sigma_0^{1/3}\equiv F_0^{2/3}B_0^{1/3}=245(8)$~MeV ~\cite{MILC:2009ltw}, $m_u=2.27(9)$~MeV, $m_d=4.67(9)$~MeV~\cite{Fodor:2016bgu}, and $m_s/\hat{m}=27.42(12)$~\cite{Blum:2014tka,Durr:2010vn,Durr:2010aw,MILC:2009ltw,Bruno:2019vup}.
This gives $\delta_\pi\approx 1\times 10^{-5}$, a prediction that can be used to benchmark our new method and its  implementation in lattice QCD.

\section{Conclusion}

We summarize our work as follows. In order to investigate the 
size of
%potentially non-negligible 
strong ISB corrections 
%at $10^{-4}$
in the free neutron decay which could affect the precise determination of $V_{ud}$, we derived an elegant representation of the leading ISB corrections to hadronic Fermi matrix elements, 
that relies on  quantum mechanical perturbation theory and the Wigner-Eckart theorem. Our result is 
consistent with the BSAG theorem~\cite{Behrends:1960nf,Ademollo:1964sr}. 
The derived correction is 
particularly suitable for implementation on the lattice,  as it requires a four-point correlation function involving readily-computable quark contraction diagrams,
and only a 20\% theoretical precision is needed to have an impact on phenomenological applications. Furthermore, one can also use the ChPT prediction in the pion sector to benchmark the lattice accuracy. Finally, one could imagine a generalization of the proposed strategy that offers another pathway to study the $K_0\rightarrow \pi_-$ transition form factor $f_+^{K\pi}(0)$ in addition to the existing methods. This could improve  the extraction of $V_{us}$ from semileptonic kaon decays.

\begin{acknowledgments}

The work of C.-Y.S. is supported in
part by the U.S. Department of Energy (DOE), Office of Science, Office of Nuclear Physics, under the FRIB Theory Alliance award DE-SC0013617, and by the DOE grant DE-FG02-97ER41014.
V.C.  acknowledges support by the U.S. DOE under Grant No. DE-FG02-00ER41132. 
X.F. is supported in part by NSFC of China under Grants No. 12125501, No. 12070131001, and No. 12141501, and National Key Research and Development Program of China under No. 2020YFA0406400.
L.J. is supported by US DOE grant DE-SC0010339 and US DOE Office of Science Early Career Award DE-SC0021147.
The work of M.G. is supported in part by EU Horizon 2020 research and innovation programme, STRONG-2020 project
under grant agreement No 824093, and by the Deutsche Forschungsgemeinschaft (DFG) under the grant agreement GO 2604/3-1.
GAM is partially supported by  U. S. Department of Energy Office of Science, Office of Nuclear Physics under Award Number DE-FG02-97ER-41014. We acknowledge partial support from the DOE Topical Collaboration ``Nuclear Theory for New Physics'', award No. DE-SC0023663.

\end{acknowledgments}

\bibliography{ISB_ref}

\end{document}